\documentstyle[graphics,sprocl,epsfig]{article}

\def\Journal#1#2#3#4{{#1} {\bf #2}, #3 (#4)}

\def\NPB{{\em Nucl. Phys.} B}
\def\NPA{{\em Nucl. Phys.} A}
\def\PLB{{\em Phys. Lett.}  B}
\def\PRL{\em Phys. Rev. Lett.}
\def\PRC{{\em Phys. Rev.} C}
\def\PRD{{\em Phys. Rev.} D}
\def\ZPA{{\em Z. Phys.} A}

\def\be{\begin{equation}}
\def\ee{\end{equation}}
\def\bea{\begin{eqnarray}}
\def\eea{\end{eqnarray}}

\begin{document}

\title{DILEPTON PRODUCTION IN ULTRARELATIVISTIC \\ HEAVY ION 
COLLISIONS\footnote{Talk given at the 30th. International Symposium on
Multiparticle Dynamics, October 9 - 15, 2000, Tihany, Hungary. } }

\author{Charles Gale}

\address{Department of Physics, McGill University\\ 3600 University St., 
Montreal, QC, H3A 2T8, Canada\\E-mail: gale@physics.mcgill.ca} 


\maketitle\abstracts{ We review the production of lepton pairs during
high energy nuclear collisions. We highlight the information being
carried out of the hot and dense strongly interacting medium.
We describe the
phenomenon of scalar-vector mixing that can take place in a dense
medium and suggest possible measurable signatures of this effect in
high energy heavy ion collisions.}

\section{Introduction}
Many probes have been proposed to map out the behaviour of hot and dense 
hadronic matter and also to highlight its eventual transition to a
plasma of quarks and gluons. It is outside the scope of this talk 
to review them
all or even partially: we will concentrate on electromagnetic 
radiation and more
specifically on the emission of thermal lepton pairs. Those suffer
minimal final state interactions and will thus carry valuable
information about their emission site. Lepton pairs should then tell
us about in-medium temperatures and densities and will constitute
information that is complementary to that carried by hadronic
observables. The rate for dilepton emission can be
expressed in terms of the finite-temperature retarded photon
self-energy \cite{McT,Weldon,GK}:
\begin{eqnarray}E_+ E_- {{d^6 R}\over{d^3 p_+ d^3 p_-}}\ &=& \ {{2
e^2}\over{(2 \pi)^6}} {{1}\over{k^4}} \, \left[ p^\mu_+ p^\nu_- +
p_+^\nu p_-^\nu - g^{\mu \nu} p_+ \cdot p_- \right]\nonumber 
\\ & \times  & {\rm Im}
\Pi_{\mu \nu}^{\rm R} (k) {{1}\over{e^{\beta \omega} - 1}}\ ,
\label{eq1}
\end{eqnarray}
where $k = p_+ + p_-$, $\omega = \sqrt{\vec{k}^2 + M^2}$, and $M^2$ =
$k^2$.  Vector Meson Dominance (VMD) couples the photon
field (real or virtual) to hadronic matter through vector mesons
\cite{VMD}. This ensures that the current-current correlator in Eq.
(\ref{eq1}) will yield to the in-medium vector meson propagator
\cite{GK}. Thus, the measurement of dilepton spectra  is a valuable
means of revealing the in-medium behaviour of vector mesons. 
We will very briefly
review the current status of the understanding of low and intermediate 
invariant mass lepton pair measurements.
Finally, we discuss the
possibility of scalar-vector mixing in hot and dense hadronic matter
and its manifestation in the lepton pair spectrum. 

\section{The low invariant mass sector}

A considerable amount of theoretical activity devoted to calculations 
in the low invariant
mass region of the dilepton spectrum has been generated by measurements
made by the CERES (NA45) collaboration at the CERN-SPS~\cite{lenkeit}. 
By first considering hadronic sources that completely explain the
proton-nucleus data, observations established that an excess over 
those sources is observed in semi-central heavy ion collisions. More
specifically, the enhancement factor defined by $N^{e^+ e^-}_{\rm
meas.} / N^{e^+ e^-}_{\rm had.\ sources}$ within the invariant
mass range $0.25 < M < 0.7$~GeV/c$^2$ is 2.6~$\pm$~0.5~(stat.)~$\pm$
0.6~(syst.), for Pb-Au collisions with $\sigma_{\rm trig.} /
\sigma_{\rm tot.}$~=~30\%.  

Theoretical interpretations of the experimental data have first involved
using a dropping (as a function of temperature and baryon density) 
vector meson mass which results from using an effective  Lagrangian
constructed to reflect the scale-invariance of the QCD action in the
massless limit \cite{BR}. It is fair to say that this approach is not
free from controversy as it implies phenomenology that is not supported 
by many-body analyses \cite{RG99} nor by QCD sum rule techniques
\cite{klingl}. Then,  
calculations of the in-medium vector meson spectral density have been done by
considering that the vector meson can couple to baryonic resonances by
interacting with nucleons, and to mesons via meson interaction.  This
will effectively broaden the $\rho$-meson spectral density and this
will in turn enhance the low mass lepton pair spectrum \cite{RW}. 
It is also natural to ask whether a cascade-type approach, representing
an incoherent sum of tree-level processes could also explain the heavy
ion dilepton data. This avenue has been explored \cite{ks}.  
The results of the theoretical calculations described above are shown
to agree well with the experimental data in Ref. [11].


The elements that are needed to make progress and to bring this issue
to some closure include 
high resolution measurements at the vector meson peaks \cite{RG99}, and
clarifying the role of baryons in the dilepton spectrum \cite{baryon}.
The relevant experimental events are either being analyzed or will hopefully 
be dealt with in an upcoming low-energy CERN run.

\section{The intermediate invariant mass sector}

\begin{figure}[ht]
\begin{center}
\includegraphics[width=8.5cm,angle=-90]{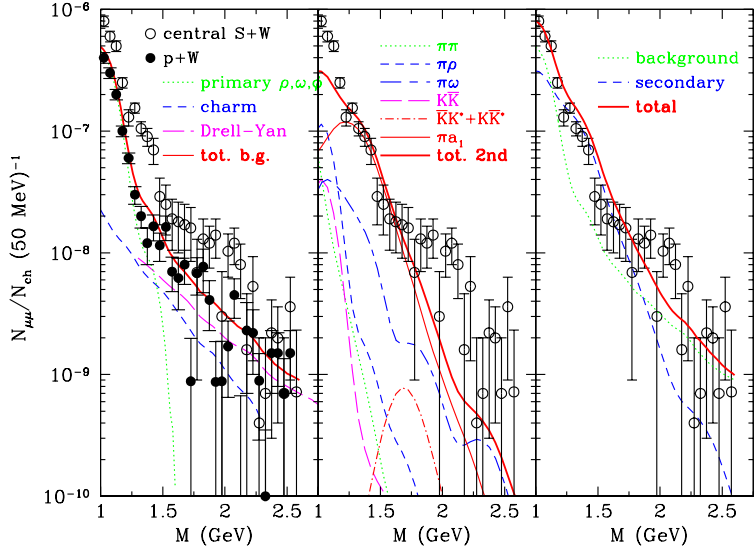}
\end{center}
\caption{Left panel: comparison of background, Drell-Yan,  and open charm
decays with p+W and S+W Helios-3 dimuon data. Middle panel: some 
mesonic reactions 
contributing to lepton pair final states are shown. Right panel: the
sum, the background, and the secondary contributions are shown with the
data \protect\cite{angelis} from central  S+W collisions.  
                \label{fig2}}
\end{figure}
Other interesting dilepton data that have displayed an enhanced yield
over that measured in proton-nucleus reactions are the dimuon spectra in
the intermediate mass region from about 1 GeV to roughly 2.5 GeV,
measured by the Helios-3 \cite{helios} and NA38/NA50 \cite{NA50} 
collaborations. Even considering sources that are known to be important
for higher masses such as Drell-Yan and correlated open charm
semileptonic decay, the experimental heavy ion data was still
underestimated.  A compelling question to ask is then whether there
exists mesonic channels similar to the ones at play for lower
invariant masses that could contribute in this regime. The first 
theoretical study in this invariant mass range concluded
that there existed such contributions \cite{GL98}.  Note however that 
studies in this region can  
generate complications that are absent for the lower masses. For
one, the usual VMD form factors will be far off-shell and furthermore,
additional couplings to high mass vector mesons is kinematically 
possible. Fortunately,
there exists a wealth of data concerning $e^+ e^- \to$ hadrons
\cite{ee} in the  appropriate invariant mass range. Those can be 
analyzed channel-by-channel and inverted by
detailed balance to build a mesonic reaction database for dilepton
production \cite{ioulia}.  Those contributions can then be used in a transport
approach to generate yields that can be directly compared with
experimental measurements. The procedure described here was used in
conjunction with a cascade-type model to compare with the Helios-3
measurements \cite{GL98}. The results are shown in Fig.~\ref{fig2}. 
One can see that the intermediate mass data are indeed reproduced once
the ``thermal'' channels are included. The measured observables call
for no additional sources.

\begin{figure}[h]
\begin{center}
\includegraphics[width=5.5cm,angle=90]{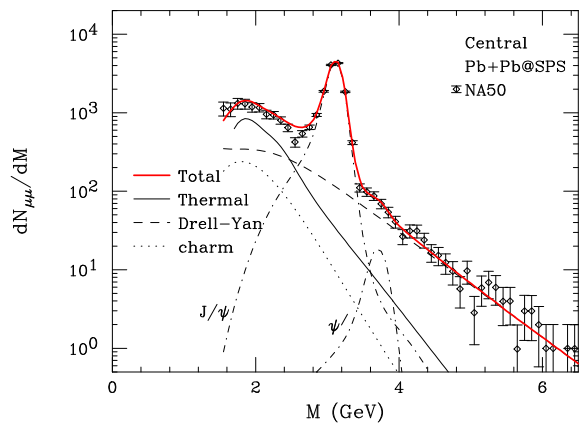}
\end{center}
\caption{Comparison with the NA50 intermediate mass data. The Drell-Yan
and open charm contributions are shown, together with a thermal
component. This contains contribution from a finite temperature meson
gas, as well as that from a QCD plasma, assuming a first-order phase
transition. The parameters of the model and details of the calculation 
will appear elsewhere \protect\cite{kgss}.} 
\label{fig3}
\end{figure}

Recently, the NA50 intermediate mass excess was analyzed in the same
spirit as that in the presentation above \cite{RS}. There also, the
thermal processes sufficed to explain the intermediate mass enhancement.
Finally, 
an updated version of the mesonic reaction database discussed earlier was
used with a hydrodynamic approach to also interpret the NA50 data
\cite{kgss}. Some preliminary results are show in Fig.~\ref{fig3}. 
The thermal characteristics of electromagnetic sources appear as 
an ubiquitous feature in nuclear collisions \cite{kamp}. 

\section{Scalar-vector mixing}

\begin{figure}[h]
\begin{flushleft}
\includegraphics[width=5cm,angle=0]{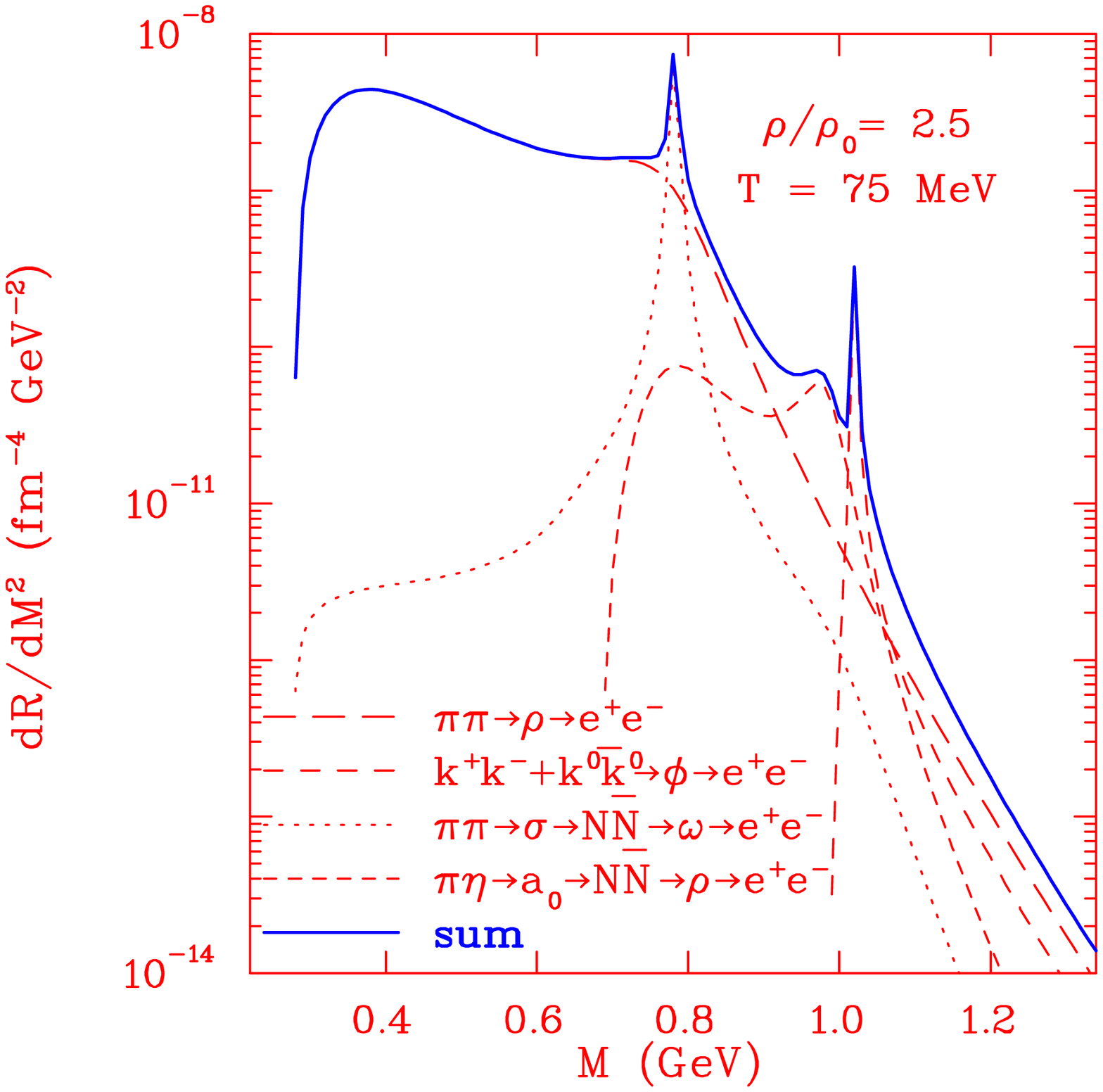}
\end{flushleft}
\vspace*{-5cm}
\begin{flushright}
\includegraphics[width=5cm,angle=0]{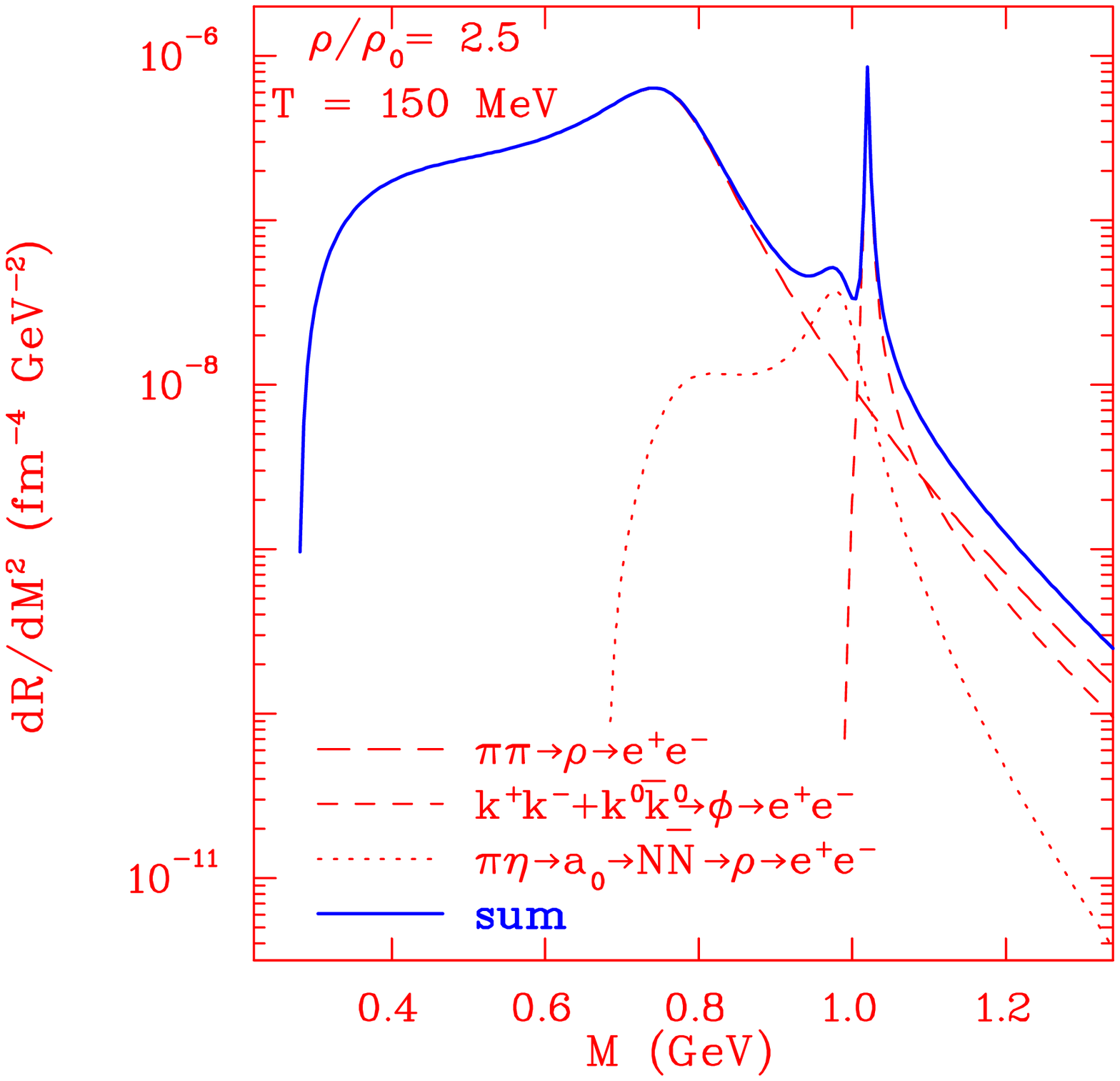}
\end{flushright}
\caption{Dilepton production rates with meson mixing effects. Left
panel: the rates from $\sigma - \omega$ and $\rho - a_0$ mixing are
shown together with those from $p$-wave $\pi \pi$ and $K \bar{K}$
annihilation. Right panel: only the rate from the $\rho - a_0$ channel
is shown together with the background processes. One observes
that the signature of this mixing channel persists up to high
temperatures, provided there exists a baryon density. } 
\label{fig4}
\end{figure}

In a finite-temperature medium, Lorentz symmetry is broken and this
will permit phenomena that are forbidden in vacuum to happen. This
breaking of Lorentz symmetry will manifest itself in longitudinal and
transverse modes having different dispersion relations, for example
\cite{GK}. At finite baryon density, scalar-vector mixing can thus
occur, more specifically $\sigma - \omega$ \cite{wfs} and $\rho - a_0$
\cite{tdg}.  The medium admits a mixed correlator: in effect a
non-diagonal self-energy. It will open new channels like $\pi \pi \to
e^+ e^-$, where the $\pi$'s annihilate into an $s$-wave, and $\pi \eta
\to e^+ e^-$. Those reactions represent genuine in-medium effects. 
It is instructive to evaluate the emission of dileptons with those mixing
effects included. Some differential rates are shown in 
Fig.~\ref{fig4}. Except for the
narrow $\omega$, most of the $\sigma - \omega$ mixing effect lies below
the $p$-wave $\pi - \pi$ channel. Perhaps a more promising candidate for
detection is connected with the appearance of a $a_0$ peak. It is
within reach  of 
high-precision measurement, such as those done by the upgraded CERES or by
HADES \cite{hades} at the GSI.

\vspace*{-0.5cm}
\section{Conclusion}
It is clear that electromagnetic radiation carries information about
the strongly interacting system that is complementary to that in
hadronic probes. The low mass sector measurements are suggestive of 
considerable
many-body effects. Similarly, high precision data  around the
vector meson peaks could highlight the signature of meson mixing. At
intermediate invariant masses, it appears that a thermal source whose
components are known is
sufficient to explain the observed enhancement. 

In summary, current
measurements have yielded exciting results and the future appears
especially promising.

\section*{Acknowledgments}
It is a pleasure to thank my collaborators. 
This work was supported in part by the Natural Sciences and Engineering
Research Council of Canada, and in part by the Fonds FCAR of the Quebec
Government.

\end{document}